\documentstyle[12pt,epsf]{ichep}
\begin{document}

\title{LATTICE GAUGE THEORY : STATUS REPORT 1994*}

\author{Junko Shigemitsu}

\affil{Physics Department, \\
The Ohio State University,\\
Columbus, Ohio 43210, USA}


\abstract{We report on recent developments and results from 
Lattice Gauge Theory.  These include determinations of the strong 
coupling constant $\alpha_s$ and the b-quark mass $M_b$ and 
an update on the Kaon 
B parameter $B_K$.  Lattice results for heavy meson decay constants
 and  semileptonic 
B and D meson decay form factors are reviewed.  The first attempts to
study the Isgur-Wise function and the radiative Penguin decay 
process, $B \rightarrow K^*\,\gamma$, on the lattice are discussed.
}

\twocolumn[\maketitle]

\fnm{7}{Plenary Talk presented at ICHEP94, Glasgow, July 1994.}

\section{Introduction}

Lattice Gauge Theory is emerging as a powerful and effective 
approach to Particle Physics.  It allows for theoretical investigations of the
 Standard Model,  that go beyond perturbation theory,
and is a practical tool for carrying out much needed calculations. A major 
part of the effort to date has been on nonperturbative studies of QCD and 
on calculations of hadronic matrix elements.  These latter quantities are 
necessary in order to compare theory with experiment, and to extract
 Standard Model parameters.

It has taken many many years and the dedicated work of a large number 
of physicists to reach this point, 
but in recent years some of the lattice results are starting to be of 
direct relevance to experimentalists and phenomenologists.
There is growing optimism that the impact of lattice gauge theory on 
phenomenology will continue to increase.
Part of this development is due to the availability of more and more 
powerful computers and sophisticated algorithms.  Just as important, 
however is the fact that lattice gauge theorists are becoming ``smarter'' 
in the types of questions they are asking and 
 how they go about systematically improving their 
calculations.  They are taking a harder look at systematic errors 
and at how to correct for them.  
There is also now a better understanding of how to match 
 from a lattice calculation to continuum physics. 

In this talk I will be focusing on 
results of lattice calculations rather than on the theory behind those 
calculations.  Hence I will not be discussing some of the nice theoretical 
developments taking place that is helping to control lattice 
calculations and that will allow for significant improvements in the 
future.  Let me just mention that 
 some of this work goes under the name of ``improved actions'' 
[1-7] 
and there has been a rebirth of sorts in this area.  Another theoretical 
development of the last couple of years that has had a huge impact on the 
field, is the revamping of lattice perturbation theory by 
Lepage and Mackenzie
\cite{pert1}.  These authors have demonstrated that their new 
improved perturbation theory is applicable for couplings at  
which numerical simulations are currently been done.  Hence it is 
possible to exploit fruitful interplays between perturbative calculations and 
nonperturbative simulation results to extract good physics.  One is also 
in a better position to justify using perturbation theory to match between 
lattice and continuum operators, as one needs to do in hadronic 
matrix element calculations.  Perturbation theory will also play a 
crucial role in the ``improvement program'', and help eliminate lattice 
artifacts. All this has contributed to strengthening 
confidence in lattice numbers.

\vspace{.1in}
\noindent
Let me conclude this Introduction by emphasizing three points,

\vspace{.1in}
\noindent
$I. \;\;\;$ There now exist results from lattice QCD, in which 
 the bulk of errors are under good control and an attempt 
has been made to correct for them.  In this category one would 
include recent calculations in,

\begin{quote}
 Light Hadron Spectroscopy

 Heavy Quark Spectroscopy ($c\bar{c}$ , $b\bar{b}$)

 Fundamental Parameters of QCD: $\alpha_s$, $M_Q$

 $B_K$

\end{quote}

\noindent
The total error is less than  10\% in many instances.  In the case of the 
important fundamental parameters, $\alpha_s$ and the b-quark mass $M_b$, 
the accuracy is considerably better.

\vspace{.1in}
\noindent
$II. \;\;$  Lattice calculations are starting to provide phenomenologically
 important hadronic matrix elements.
\begin{quote}
 Meson Decay Constants : $f_D$, $f_{D_s}$, $f_B$, $f_{B_s}$,
 $f_B \sqrt{B_B}$,..

 Semileptonic Decays : $D \rightarrow K (K^*)$ $l \nu$

 Isgur-Wise Function

 Rare B Decays :
 $B \rightarrow K^* \,\gamma$
\end{quote}
\noindent
Current lattice results for these matrix elements have uncertainties that are 
generally larger than 10\% and one is still exploring better ways to 
approach the problem and reduce statistical and systematic errors.  However, 
prospects are good that in due  time accurate results for 
quantities such as $f_B\,\sqrt{B_B}$ will become available.  

\vspace{.1in}
\noindent
$III. \;$ The Lattice continues to be a versatile tool to investigate
Standard Model and Quantum Field Theory issues.  It provides a 
general purpose, mathematically well defined regularization scheme for 
quantum field theories.
  A partial list of topics under study using 
lattice techniques includes,

\begin{quote}
  QCD at finite T

  Structure of Hadrons

  Test of the Parton Model

  Electroweak phase transition

  Confinement Mechanism,  Monopoles

  Strong CP

  Higgs and Fermion Mass Upper Bounds

  Chiral Gauge Theories

  Quantum Gravity

\end{quote}

\vspace{.1in}
\noindent
Due to time constraints this talk will focus only on the first two 
category of topics.  Interesting parallel session talks were presented 
on several of the items in the third category\cite{other}.  Apologies to those 
speakers for not covering their work.  Readers are urged to read their 
contributions to these proceedings.

\vspace{.3in}

\section{Light Hadron Spectroscopy}

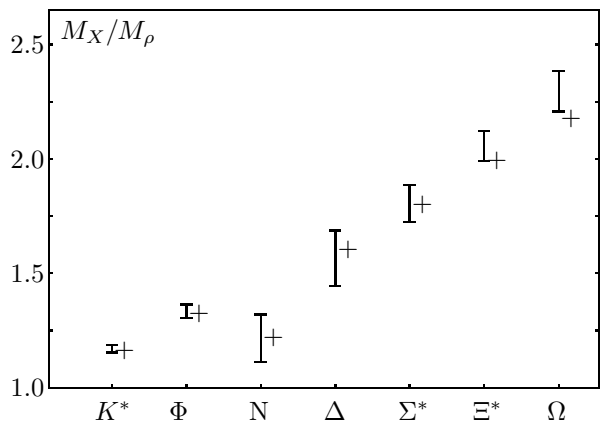
\begin{figure}[htb]
\begin{center}
\setlength{\unitlength}{.006in}
\begin{picture}(500,400)(0,0)
\put(10,50){\line(0,1){330}}
\multiput(10,100)(0,50){6}{\line(1,0){3}}
\multiput(487,100)(0,50){6}{\line(1,0){3}}
\put(6,50){\makebox(0,0)[r]{1.0}}
\put(6,150){\makebox(0,0)[r]{1.5}}
\put(6,250){\makebox(0,0)[r]{2.0}}
\put(6,350){\makebox(0,0)[r]{2.5}}

\put(20,360){\makebox(0,0)[l]{$M_X / M_\rho$}}

\put(10,50){\line(1,0){480}}
\put(10,380){\line(1,0){480}}
\put(490,50){\line(0,1){330}}

\multiput(65,50)(65,0){8}{\line(0,1){3}}
\put(50,30){\makebox(0,0)[l]{$K^*$}}
\put(115,30){\makebox(0,0)[l]{$\Phi$}}
\put(185,30){\makebox(0,0)[l]{N}}
\put(250,30){\makebox(0,0)[l]{$\Delta$}}
\put(315,30){\makebox(0,0)[l]{$\Sigma^*$}}
\put(380,30){\makebox(0,0)[l]{$\Xi^*$}}
\put(445,30){\makebox(0,0)[l]{$\Omega$}}

\put(85,82.8){\makebox(0,0)[r]{+}}
\put(65,84){\line(0,1){3.2}}
\put(65,84){\line(0,-1){4.2}}
\put(60,87.2){\line(1,0){10}}
\put(60,80.8){\line(1,0){10}}

\put(150,115.4){\makebox(0,0)[r]{+}}
\put(130,117){\line(0,1){6}}
\put(130,117){\line(0,-1){6}}
\put(125,123){\line(1,0){10}}
\put(125,111){\line(1,0){10}}

\put(215,94.4){\makebox(0,0)[r]{+}}
\put(195,93.2){\line(0,1){20.8}}
\put(195,93.2){\line(0,-1){20.8}}
\put(190,114){\line(1,0){10}}
\put(190,72.4){\line(1,0){10}}

\put(280,170.8){\makebox(0,0)[r]{+}}
\put(260,163){\line(0,1){24.4}}
\put(260,163){\line(0,-1){24.4}}
\put(255,187.4){\line(1,0){10}}
\put(255,138.6){\line(1,0){10}}

\put(345,210.6){\makebox(0,0)[r]{+}}
\put(325,211.2){\line(0,1){16}}
\put(325,211.2){\line(0,-1){16}}
\put(320,227){\line(1,0){10}}
\put(320,195){\line(1,0){10}}

\put(410,249.2){\makebox(0,0)[r]{+}}
\put(390,261){\line(0,1){13}}
\put(390,261){\line(0,-1){13}}
\put(385,274){\line(1,0){10}}
\put(385,248){\line(1,0){10}}

\put(475,285.4){\makebox(0,0)[r]{+}}
\put(455,309.2){\line(0,1){17.8}}
\put(455,309.2){\line(0,-1){17.8}}
\put(450,327){\line(1,0){10}}
\put(450,291.4){\line(1,0){10}}

\end{picture}
\end{center}
\label{hadspec}
\caption{Light hadron spectrum results from the IBM GF11 group[9].
Masses are measured in terms of the $\rho$ meson mass, $M_\rho$.
Experimental numbers are denoted by crosses.}
\end{figure}

Testing whether QCD, our theory of the strong interactions,  contains 
a low lying spectrum that is consistent with observed light hadrons remains 
a basic, fundamental problem in nonperturbative QCD.  
Fig.1. shows results by the IBM GF11 group for a first principles QCD
calculation of light hadron masses\cite{lightsp}. 
  Plotted are hadron masses measured in 
units of the $\rho$ meson mass $M_\rho$.  Three parameters of the theory, 
the inverse lattice spacing, $a^{-1}$, and two quark masses, $M_u = M_d$ and 
$M_s$,  need to be fixed by experiment.  The GF11 group uses the $\rho$ meson 
mass, $M_\rho$ to fix
 $a^{-1}$. 
$M_u$ and $M_s$ are determined by the pion and kaon masses respectively.
  Having done so, 
there are no adjustable paramenters left and the remaining hadron mass 
values become predictions of the theory.  
The GF11 group has attempted to correct for three out of the four main 
sources of systematic errors that contaminate many lattice calculations.
The three are finite volume errors, finite lattice spacing errors and 
quark mass errors (the fact that present technology forces people to work 
with quark masses larger than up- and down-quark masses). 
  The fourth common error comes from the 
neglect of light quark vacuum polarization contributions (the ``quenched 
approximation'').  The results of Fig.1. have not been corrected
 for quenching errors.    Nevertheless one sees that, agreement between 
lattice results and the real world is good.  These calculations 
can and will be improved upon in the future.  Eventually calculations 
going beyond the quenched approximation will become feasable.  This is 
 on the one hand a  
question of algorithmic breakthroughs, or failing that the availability of 
ever more powerful computers. On the other hand 
 progress in dealing with improved actions,  should allow lattice gauge 
theorists to work on smaller lattices and with larger lattice spacings 
than many people currently anticipate.  So the prospect for having 
high statistics unquenched results on realistic systems may not be too 
far off in the future.

\vspace{.3in}

\section{Quarkonium Physics}

Systems involving one or more heavy quarks and/or antiquarks have been 
the focus of intense investigations on the lattice in recent years.  In this 
section I discuss heavy-heavy (quarkonium) systems and will return to 
heavy-light systems in later sections.  
At first sight one may worry about working with quarks with masses 
larger than the inverse lattice spacing,
$M_Q \geq a^{-1}$.  Several ways to deal with heavy quarks have been 
developed in recent years and demonstrate that this does not pose a problem. 
For instance, one has :

\begin{itemize}
\item The Static Approach\cite{static} \\
      Here one works in the $M_Q \rightarrow \infty$ limit. This approach 
      is being applied extensively to 
      heavy-light systems but is not appropriate for 
      heavy-heavy systems.

\item The Non-Relativistic Approach (NRQCD) \\
\cite{nrqcd1,nrqcd2,nrqcd3,nrqcd4} \\
      One works with heavy fermions with a nonrelativistic dispersion 
      relation

$$ E(\vec{p}) = M_1 + {\vec{p}^2 \over 2M_2} + \;\; ....  $$

     The rest mass $M_1$ is eliminated as a dynamical variable and one 
     is working with an expansion in $v^2$.  For $\Upsilon$ $v^2$ is 
     typically of order $\sim 0.1$ and for charmonium one has 
     $v^2 \sim 0.3$.

\item The Heavy Wilson Approach\cite{hwils1,hwils2} \\
      In this approach a generalized set of operators is used for the 
      fermionic action, the coefficients of which are kept to 
      all orders in the quark mass.  This action encompasses as special 
      cases the Wilson and the non-relativistic action.  The only expansion 
      parameters left are $pa$ and the gauge coupling.

\end{itemize}

\begin{figure}[htb]
\begin{center}
\setlength{\unitlength}{.02in}
\begin{picture}(130,120)(10,930)
\put(15,935){\line(0,1){120}}
\put(140,935){\line(0,1){120}}
\multiput(13,950)(0,50){3}{\line(1,0){4}}
\multiput(14,950)(0,10){10}{\line(1,0){2}}
\put(12,950){\makebox(0,0)[r]{9.5}}
\put(12,1000){\makebox(0,0)[r]{10.0}}
\put(12,1050){\makebox(0,0)[r]{10.5}}
\put(12,1057){\makebox(0,0)[r]{GeV}}
\put(15,935){\line(1,0){125}}
\put(15,1055){\line(1,0){125}}



\put(27,930){\makebox(0,0)[t]{${^1S}_0$}}
\put(25,943){\circle*{3}}
\put(30,942){\circle{3}}

\put(52,930){\makebox(0,0)[t]{${^3S}_1$}}
\multiput(43,946)(3,0){7}{\line(1,0){2}}
\put(50,946){\circle*{3}}
\put(55,946){\circle{3}}
\put(60,946){\makebox(0,0){$\Box$}}

\multiput(43,1002)(3,0){7}{\line(1,0){2}}
\put(50,1004){\circle*{3}}
\put(50,1004){\line(0,1){3}}
\put(50,1004){\line(0,-1){3}}
\put(55,1003){\circle{3}}
\put(55,1004){\line(0,1){2}}
\put(55,1002){\line(0,-1){2}}
\put(60,1009){\makebox(0,0){$\Box$}}
\put(60,1011){\line(0,1){7}}
\put(60,1008){\line(0,-1){7}}

\multiput(43,1036)(3,0){7}{\line(1,0){2}}
\put(50,1034){\circle*{3}}
\put(50,1034){\line(0,1){12}}
\put(50,1034){\line(0,-1){12}}
\put(55,1029){\circle{3}}
\put(55,1030){\line(0,1){13}}
\put(55,1028){\line(0,-1){13}}

\put(92,930){\makebox(0,0)[t]{${^1P}_1$}}

\multiput(83,990)(3,0){7}{\line(1,0){2}}
\put(90,987){\circle*{3}}
\put(90,987){\line(0,1){2}}
\put(90,987){\line(0,-1){2}}
\put(95,990){\circle{3}}
\put(100,987){\makebox(0,0){$\Box$}}
\put(100,989){\line(0,1){3}}
\put(100,986){\line(0,-1){3}}

\multiput(83,1026)(3,0){7}{\line(1,0){2}}
\put(90,1032){\circle*{3}}
\put(90,1032){\line(0,1){7}}
\put(90,1032){\line(0,-1){7}}
\put(95,1025){\circle{3}}
\put(95,1026){\line(0,1){4}}
\put(95,1024){\line(0,-1){4}}
\put(100,1024){\makebox(0,0){$\Box$}}
\put(100,1026){\line(0,1){4}}
\put(100,1023){\line(0,-1){4}}

\put(120,930){\makebox(0,0)[t]{${^1D}_2$}}
\put(120,1020){\circle*{3}}
\put(120,1020){\line(0,1){7}}
\put(120,1020){\line(0,-1){7}}
\put(123,1031){\makebox(0,0){$\Box$}}
\put(123,1033){\line(0,1){6}}
\put(123,1030){\line(0,-1){6}}

\end{picture}
\end{center}
\caption{$\Upsilon$ spectrum results from (a) NRQCD ($n_f=0$): filled circles,
(b) NRQCD ($n_f=2$): open circles
 and (c) UK-NRQCD ($n_f=0$): boxes.  The 
dashed horizontal lines correspond to experiment. For the P states 
experimental numbers for the center of mass of triplet P states are shown.}
\label{upsfig}
\end{figure}
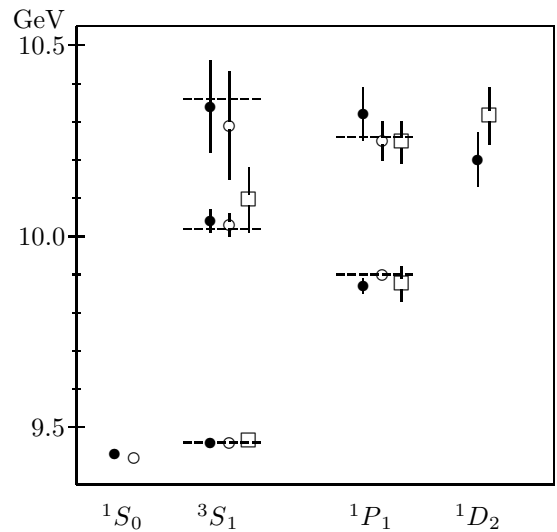

\noindent
 There are several reasons why 
studies of quarkonia on the lattice are particularly rewarding.
\begin{enumerate}
\item There exists a wealth of experimental data on these systems and more 
continues to be accumulated. Quarkonia provide good opportunities to test 
lattice methods and also for the lattice to present experimentalists with 
predictions.  

\item  Studies of quarkonia are computationally relatively cheap.  High 
statistics results are now available.  They 
 demonstrate what can be achieved when 
calculations are not statistics limited, so that one can concentrate on 
systematics.  

\item Systematic effects are easier to understand and control.  The NRQCD 
  collaboration, for instance,
\cite{nrqcd2,nrqcdsp} includes relativistic corrections through 
  {\it O}($M_Qv^4$) and finite lattice spacing corrections in their quark 
  action through {\it O}($a^2$).  Finite volume effects are negligible for 
  these ``small'' mesons.  Corrections to counteract 
  quenching errors can often be 
  estimated.  Furthermore calculations with $n_f=2$ light dynamical 
  fermions have now appeared and to date there have been no surprises.  
  Unquenching effects have been observed where expected,  and have been 
  demonstrated to be small where they were predicted to be small.

\item Studies of quarkonia on the lattice provide the 
      basis for determining fundamental parameters of QCD such as 
      $\alpha_s$ and heavy quark masses.  These important results 
      are discussed in the next section.  

\item Experience gained here will be crucial in studies of heavy-light 
      systems, such as the $B$ and $D$ meson systems.  

\end{enumerate}

\begin{figure}[htb]
\begin{center}
\setlength{\unitlength}{.02in}
\begin{picture}(200,140)(5,280)
\put(25,285){\line(0,1){125}}
\put(155,285){\line(0,1){125}}
\multiput(23,300)(0,50){3}{\line(1,0){4}}
\multiput(24,290)(0,10){10}{\line(1,0){2}}
\put(22,300){\makebox(0,0)[r]{3.0}}
\put(22,350){\makebox(0,0)[r]{3.5}}
\put(22,400){\makebox(0,0)[r]{4.0}}
\put(22,410){\makebox(0,0)[r]{GeV}}
\put(25,285){\line(1,0){130}}
\put(25,410){\line(1,0){130}}

\put(80,400){\makebox(0,0)[l]{$\;${\bf (Preliminary)}}}

     \put(40,280){\makebox(0,0)[t]{$\eta_c$}}
\multiput(33,298)(3,0){5}{\line(1,0){2}}
     \put(38,300){\circle*{3}}
     \put(42,300.5){\makebox(0,0){$\Box$}}


     \put(60,280){\makebox(0,0)[t]{$J/\Psi$}}
\multiput(53,310)(3,0){5}{\line(1,0){2}}
     \put(58,309){\circle*{3}}
     \put(62,308.5){\makebox(0,0){$\Box$}}
\multiput(53,369)(3,0){5}{\line(1,0){2}}

\multiput(53,369)(3,0){5}{\line(1,0){2}}
     \put(58,373){\circle*{3}}
     \put(58,373){\line(0,1){9}}
     \put(58,373){\line(0,-1){9}}
     \put(62,372){\makebox(0,0){$\Box$}}
     \put(62,374){\line(0,1){18}}
     \put(62,370.5){\line(0,-1){18}}

     \put(80,280){\makebox(0,0)[t]{$h_c$}}
\multiput(73,353)(3,0){5}{\line(1,0){2}}
     \put(78,353){\circle*{3}}
     \put(82,352.5){\makebox(0,0){$\Box$}}

     \put(100,280){\makebox(0,0)[t]{$\chi_{c0}$}}
\multiput(93,342)(3,0){5}{\line(1,0){2}}
     \put(98,346){\circle*{3}}
     \put(102,345.5){\makebox(0,0){$\Box$}}
     \put(102,347.5){\line(0,1){1}}
     \put(102,344.5){\line(0,-1){1}}

     \put(115,280){\makebox(0,0)[t]{$\chi_{c1}$}}
\multiput(108,351)(3,0){5}{\line(1,0){2}}
     \put(113,350){\circle*{3}}
     \put(117,348.5){\makebox(0,0){$\Box$}}
     \put(117,350.5) {\line(0,1){2}}
     \put(117,347.5) {\line(0,-1){2}}

     \put(130,280){\makebox(0,0)[t]{$\chi_{c2}$}}
\multiput(123,356)(3,0){5}{\line(1,0){2}}
     \put(130,356.6){\circle*{3}}

%


\end{picture}
\end{center}
\label{jpsifig}
\caption{Charmonium spectrum from the (a)NRQCD: full circles
and (b)Fermilab: boxes
, collaborations. Dashed lines show experimental
 numbers.}
\end{figure}
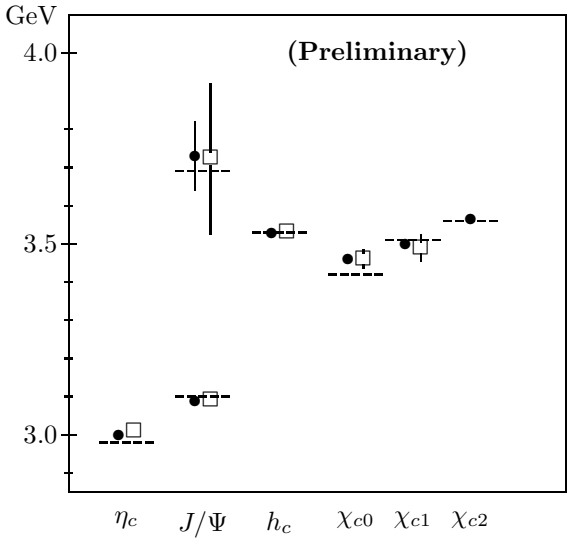

\Fref{upsfig} shows the $\Upsilon$ spectrum using non-relativistic fermions
\cite{nrqcdsp,uknrqcd}.
The NRQCD collaboration has results both from quenched configurations 
($n_f=0$) and from configurations with two light dynamical quarks ($n_f=2$).
Charmonium results from the Fermilab\cite{hwilssp} and 
NRQCD\cite{nrqcdjpsi}
 collaborations are shown in \fref{jpsifig}.  The charmonium results
were obtained in the quenched approximation.  The zero of energy was adjusted 
to give the correct $\Upsilon$ or the center of mass of the $J/\Psi - \eta_c$
states. The inverse lattice spacing was obtained from the 1S-1P splitting in 
the case of charmonium.  For $\Upsilon$ an average $a^{-1}$ from the 
1S-1P and 1S-2P splittings was used.

The general good agreement between simulation results and experiment 
gives us confidence that heavy quarks on the lattice are under good 
control.  Further improvements are also possible here.  {\it O}($M_Qv^6$) 
spin dependent terms are being included in the NRQCD action to improve 
on the splittings among the P-states.  Gauge configurations with 
{\it O}($a^2$) errors removed should also be used in the future.

\vspace{.3in}

\section{Fundamental Parameters of QCD}

One major goal of the lattice effort is to contribute 
towards nailing down fundamental parameters of the Standard Model.  There has 
been significant progress in this area.  Here we discuss fundamental 
parameters of QCD.  In later sections, 
matrix elements relevant to investigations of the electroweak theory 
will be reviewed. 
Precision calculations of Quarkonium spectra allow good 
determinations of the inverse lattice spacing, $a^{-1}$, and 
of the value of the lattice bare quark mass, $M_Q^0$.  This information 
can then be converted into determinations of $\alpha_s$ and heavy quark 
pole and $\overline{MS}$ masses.

\subsection{The Strong Coupling Constant}

\vspace{.1in}
\noindent
Lattice simulations give the 
 $\Upsilon$ 1S-1P or 1S-2S splittings measured in units of the 
inverse lattice spacing.  $a^{-1}$ is determined
for instance in GeV's by combining 
the dimensionless simulation numbers with experimental values for 
the splittings.  The actual 
value of $\alpha_s(C/a)$ (``C'' is a known constant) is obtained by 
focusing on some short distance quantity, for which both its 
nonperturbative simulation value and its perturbative expansion 
are known. A useful candidate is the ``average plaquette'', $W_{1,1}$,
whose expansion is given by\cite{logw1,logw2,pert1}, 
$$
-lnW_{1,1} = \qquad \qquad \qquad \qquad \qquad \qquad \qquad \qquad \qquad
$$
\begin{equation}
 {4\pi \over 3}\alpha_V({3.41 / a})
 (1 - \alpha_V(3.41/a)\,[1.185 + 0.070n_f]) 
\end{equation}

\noindent
$\alpha_V$ is a coupling introduced in \cite{pert3,pert1}
 and defined in terms of 
the static quark potential.  Once $\alpha_V$ at scale $3.41/a\;GeV$
is known, one can convert to other definitions such as $\alpha_{\overline
{MS}}$ and run the coupling to other scales using the two- or three-loop 
beta-function.  

\vspace{.2in}
\noindent
\underline{A.  $\alpha_s$ from Quenched Data}

\vspace{.1in}
\noindent
The first lattice determinations of $\alpha_s$ 
following the methods just described\cite{alpha1,alpha2,nrqcd3},
 used simulations 
in the quenched approximation, i.e. one initially had $\alpha_V^{(n_f=0)}
(3.41/a)$, and the effects of light quarks had to be included a posteriori. 
This was accomplished by running $\alpha_V^{(0)}$ from scale $3.41/a$ down 
to scales typical of momenta exchanged in $\Upsilon$ or $J/\Psi$ mesons.  
$\alpha_V^{(0)}$ at the lower scale 
 is equated to the coupling of an $n_f=3$ theory, $\alpha_V^{(3)}$, since it 
is at this scale that the two couplings  must produce the same physics 
such as splittings. $\alpha^{(3)}$  can then be evolved from the quarkonium 
scale through the charm and bottom thresholds up to the Z scale.  
 Based on $\Upsilon$ results from two lattice groups,
 the latest Particle Data 
Group booklet quotes a lattice result of,

\begin{equation}
\left. \alpha^{(n_f=5)}_{\overline{MS}}(M_Z)\right|_{lattice}
    =\; 0.110(6)  
\end{equation}

\vspace{.1in}
\noindent
This is an average over Fermilab's 0.108(6) and NRQCD's  0.112(4). Both 
numbers were obtained from quenched configurations, with dynamical 
light quark effects incorporated as described above.  The conversion from 
$\alpha_V$ to $\alpha_{\overline{MS}}$ is described below in eq.(\ref{alphavms}
).

\vspace{.2in}
\noindent
\underline{B.  $\alpha_s$ using Unquenched Data}

\vspace{.1in}
\noindent
Recently, lattice results for $\alpha_s$ based on calculations with 
$n_f=2$ dynamical light quarks have appeared.  The NRQCD collaboration 
extrapolates directly to $n_f=3$ at scale $3.41/a$ using their $n_f=0$ 
and $n_f=2$ data\cite{nrqcdalph}.
  They no longer need to run down to a much lower 
scale in order to match onto an $n_f=3$ theory.  NRQCD also uses 
two different splittings, the 1S-1P and 1S-2S splittings, to carry 
out two separate analyses.  From the 1S-1P splitting they find,

$$
 \alpha_V^{(n_f)}(8.2GeV) = \left\{ \begin{array}{ll}
                      0.1548(23) &  \;\; n_f = 0 \\
                      0.1800(16) &  \;\; n_f = 2
                      \end{array}
\right.
$$

\noindent
Extrapolating to $n_f = 3$ one gets,

\begin{equation}
 \alpha_V^{(3)}(8.2GeV) = 0.1959(34)  
\label{alphav}
\end{equation}
A similar analysis based on the 1S-2S splitting gives 
$ \alpha_V^{(3)}(8.2GeV) = 0.1958(46)  $.  

\vspace{.1in}
\noindent
In order to compare with other non lattice determinations of $\alpha_s$, 
one needs to convert to the commonly used $\alpha_{\overline{MS}} $
scheme.  The $\alpha_V$ and $\alpha_{\overline{MS}}$ schemes are 
related by,

\begin{equation}
  \alpha_{\overline{MS}}(Q) = \alpha_V(e^{5/6}Q) (1 + {2 \over \pi} \alpha_V
+ {\cal O}(\alpha_V^2))
\label{alphavms}
\end{equation}

\noindent
so that Eq.(\ref{alphav}) becomes,

\begin{equation}
 \alpha_{\overline{MS}}^{(3)}(3.56GeV) = 0.2203(84) 
\end{equation}

\noindent
Errors are now dominated by higher orders in the $\alpha_V$ -
$\alpha_{\overline{MS}}$
 conversion formula, rather than by $a^{-1}$ uncertainties.

\vspace{.1in}
\noindent
Evolving to the Z boson scale, one finds,

\begin{equation}
 \alpha^{(5)}_{\overline{MS}}(M_Z) = 0.115(2) 
\label{alphamz}
\end{equation}

\vspace{.2in}
\noindent
This should be compared with the world average quoted at this 
conference of $\alpha_{\overline{MS}}(M_Z) = 0.117(5)$\cite{avalpha}.  
Had one used the old  method of evolving down to a typical $\Upsilon$
scale and matching onto $n_f=3$, the resulting numbers would have been,

\begin{equation}
 \alpha^{(5)}_{\overline{MS}}(M_Z) = 0.112(4) \;\;\;
              (from\; n_f=0\; data)
\end{equation}

\begin{equation}
 \alpha^{(5)}_{\overline{MS}}(M_Z) = 0.115(3) \;\;\;
                      (from\; n_f=2\; data)
\end{equation}
i.e. the previous method gives values consistent with direct extrapolation 
to $n_f=3$, but with larger errors.  In the future one would like to 
see calculations repeated on configurations with $n_f=4$ or 
$n_f=3$ light dynamical quarks. Knowing the next order in the $\alpha_V$ - 
$\alpha_{\overline{MS}}$ conversion formula, eq.(\ref{alphavms}), 
would contribute significantly to further reduction in the errors.

\vspace{.1in}
\noindent
A recent paper describes an independent calculation of $\alpha_s$ using 
$n_f=2$ dynamical configurations\cite{alpha4}.  The authors 
employ the same methods but start from the charmonium spectrum obtained 
using Wilson fermions. 
They find $\alpha_s$ at the Z mass consistent with the above lattice 
determinations, but with larger errors.  Other lattice approaches to 
$\alpha_s$ are also being pursued\cite{alpha5,alpha6}.
  In particular, reference \cite{alpha6} has 
 accurate results for the pure SU(3) gauge theory which agrees with 
calculations by the groups already mentioned in the same limit. 
  Once the calculations of reference \cite{alpha6}
have been generalized to include quarks, one will have another 
independent first principles lattice determination of $\alpha_s$ that can 
be compared with Eq.(\ref{alphamz}).  

\subsection{ Heavy Quark Masses}

By fitting to the experimental $\Upsilon(1S)$ mass one can fix 
the value of the bare lattice mass, $M_b^0$, 
that appears in the lattice action.  
$M_b^0$ can then be converted into other definitions of quark masses, such 
as the pole mass, using perturbation theory
($M_b^{(pole)} = Z_m\,M_b^0$).  Another approach to, extracting 
the pole mass, starts from the following relation between the $\Upsilon$ mass, 
the b-quark pole mass and the binding energy inside an $\Upsilon$ meson.

\begin{equation}
M_{\Upsilon} = 2 ( M_b^{(pole)} - E_0 ) +  E_{NR}(\Upsilon)
\label{polemass}
\end{equation}
$M_{\Upsilon}$ is known from experiment,  $E_0$ can be calculated 
perturbatively and $E_{NR}$ can be extracted from simulations (the 
binding energy corresponds to $E_{NR} - 2E_0$).
This allows for another determination of $M_b^{(pole)}$. 
 The NRQCD collaboration has 
estimated $M_b^{(pole)}$ using both methods and finds that the two methods
  lead to consistent values.  They now have results based on $n_f=0$ and 
$n_f=2$ configurations\cite{nrqcdmb}. 

\vspace{.1in}

  $ M^{(pole)}_b = (5.0 \pm 0.2) GeV\;\;\; (n_f = 2) $

   $M_b^{(pole)} = (4.94 \pm 0.15) GeV \;\;\; (n_f = 0)$

\vspace{.1in}
\noindent

\vspace{.2in}
\noindent
For several applications the $\overline{MS}$ definition of a running 
quark mass is more appropriate.  The bare lattice b-quark mass can 
also be converted to a $M_b^{\overline{MS}}$ value. The two-loop formula 
relating pole and $\overline{MS}$ masses was calculated in reference 
\cite{gray} 
in terms of the $\overline{MS}$ coupling.  Converting to the $\alpha_V$ 
coupling one finds,

\begin{equation}
M_b = M_b^{\overline{MS}}(M_b)
(1 + {4 \over 3\,\pi}\alpha_V(0.22\,M_b) + 0.164\,\alpha_V^2 + ..)
\label{polems}
\end{equation}

\noindent
$M_b \equiv M_b^{(pole)}$.  
In deriving eq.(\ref{polems}), one has used the scheme of reference 
\cite{pert3} to set 
the scale.  Rather than use $\alpha$ evaluated at 
 the pole mass $M_b$ one has 
$\alpha_V(0.22\,M_b)$.  With this scheme the two-loop coefficient has 
been reduced drastically 
compared to the $\overline{MS}$ scheme of reference \cite{gray}
but the coupling is evaluated at a much more infrared scale. 
Finally one needs the relation between the bare lattice mass $M_b^0$ 
and the $\overline{MS}$ mass.  This can be done by combining eq.(\ref{polems})
 with the relation between pole mass and bare lattice mass.

\begin{equation}
 M_b = Z_m(\alpha_V(q^*_m))\,M^0_b  
\label{mbmb0}
\end{equation}
The mass renormalization $Z_m$ and the scale $q^*_m$ have been calculated 
in reference \cite{morning}. 
 From (\ref{polems}) and (\ref{mbmb0}) one derives,

\begin{equation}
    M^{\overline{MS}}_b(M_b) = M^0_b ( 1 +
0.06\,\alpha_V(q_{\overline{MS}}) + ..)
\end{equation}
\noindent
with $q_{\overline{MS}} \sim {0.8 \over a}$.

\vspace{.2in}
\noindent
Hence,
\begin{equation}
 M^{\overline{MS}}_b(M_b) \approx M^0_b = 4.0(1)GeV  
\end{equation}

\noindent
There is a  large difference between pole and $\overline{MS}$ masses, 
so it is crucial to pick the correct type of mass for each application.

\vspace{.1in}
\noindent
The Fermilab group has estimated the charm quark mass from their 
charmonium studies using similar methods\cite{fermimc}.
They find $M_c^{(pole)} \sim 1.55\,GeV$.

\vspace{.2in}
\noindent
Finally one should mention that several attempts have been made over the years 
to estimate quark masses from light quark physics.  These calculations 
suffer at the moment from unknown systematic errors.  It is not clear how 
to account for quenching effects.  One typically uses zero mass perturbative 
formulas to relate bare and running masses, which for the charm quark, 
for instance could introduce large distortions. 
( The mass determinations 
by the Fermilab and NRQCD groups based on quarkonia that were described
above are not 
plagued by these uncertainties). Recent quark mass estimates from light 
physics have appeared from the Rome group\cite{rome}.
 They find $M_c^{\overline{MS}}
(2GeV) = (1.47 \pm 0.28)GeV$ and $M_s^{\overline{MS}}(2GeV) = (127 \pm 
18)MeV$.  Another group, UKQCD
\cite{ukqcdsm}, finds  $M_s^{\overline{MS}}(2GeV) = (109 \pm 
11)MeV$.

\vspace{.3in}

\section{The $B_K$ Parameter}

Nonperturbative determination of the Kaon B-parameter,

\begin{equation}
 B_K = {{<\overline{K^0} | {\cal O}_{\Delta S = 2} | K^0>} \over
          { {8 \over 3} m^2_K f^2_K }}
\end{equation}

\vspace{.1in}
\noindent
is one of the important contributions to phenomenology from the lattice.
Major progress has taken place in recent years in reducing errors. This 
has been possible due to, 

\begin{itemize}

\item Better understanding of perturbative corrections to
      lattice operators\cite{bk2}. \\
      Use of the Lepage - Mackenzie coupling\cite{pert1}.

\item Tests demonstrating that different choices for lattice operators 
      lead to the same physical results\cite{bk1,bk3}.

\item  Control over extrapolations in ``a'' (quadratic)\cite{bk1}.

\item  First tests of effects due to dynamical quarks\cite{bk3,bk4}.

\end{itemize}

\noindent
The current best value, using the NDR scheme and with scale fixed at 2GeV 
gives\cite{bk1}, 

\begin{equation}
         B_K(NDR,2GeV) = 0.616 \pm 0.020 \pm 0.017 
\end{equation}

\noindent
It is customary to quote the scale independent quantity, 
$\hat{B}_K = \alpha_s^{-6/25}B_K$.  Using the continuum $\alpha_s$ evaluated 
at 2GeV with $\Lambda_{\overline{MS}}^{(4)} = 300MeV$, one finds ,

\begin{equation}
          \hat{B}_K = 0.825 \pm 0.027 \pm 0.023  
\label{bsubk}
\end{equation}

\vspace{.2in}
\noindent
Eq.(\ref{bsubk}) includes statistical and systematic
 errors coming from the lattice 
calculation.  There is a further systematic error coming from perturbative 
uncertainties in going to eq.(\ref{bsubk}) of about 3\%. 
The recent reduction of errors in $B_K$ significantly 
reduces the allowed region in the $\rho$ - $\eta$ plane of CKM parameters.
At the moment the lattice $B_K$ results have the smallest errors
 among theoretical estimates of this quantity. 
In the future more detailed studies of quenching effects, of the 
light quark mass dependence and of the use of degenerate quarks 
 in the above calculations,
 should further improve the lattice calculations.

\vspace{.3in}

\section{Heavy Mesons Decay Constants}
\vspace{.2in}
\noindent

A major challenge to the lattice community is to produce  
accurate values for the decay constants of D and B mesons\cite{sonimr}. 
 For instance, 
 combinations such as $f_B\,\sqrt{B_B}$ or $f_{B_s}\,\sqrt{B_{B_s}}/
f_B\,\sqrt{B_B}$ are crucial inputs into analyses of $B^0 - \overline
{B^0}$ mixing phenomena.  These nonperturbative QCD numbers are necessary 
in order to determine CKM matrix elements from experimental data.

\vspace{.2in}
\noindent
The pseudoscalar decay constant, $f_{PS}$, is defined (in Euclidean space) by,

\begin{equation}
<0|\,A_\mu\,|PS ; p> = p_\mu \, f_{PS}
\end{equation}
where $A_\mu$ is the axial vector current and $|PS;p>$ is the pseudoscalar 
state with momentum $p_\mu$.  A lattice determination of $f_{PS}$ involves 
first calculating and fitting the following correlation functions.

 $$   \sum_{\vec{x}} <0|\,A^L_4(\vec{x},t)\,{\cal P}(0)\,|0> $$
 $$ t \gg 1 \;\;\rightarrow { {<0|\,A^L_4\,|PS>\;
<PS|\,{\cal P}\,|0>} \over {2\;M_{PS}a^3}}\;e^{-M_{PS} \cdot t}$$
\begin{equation}
 \equiv \zeta_{AP} \; e^{-M_{PS} \cdot t}
\label{apcorr}
\end{equation}

\vspace{.1in}

 $$   \sum_{\vec{x}} <0|\,{\cal P}^{\dagger}(\vec{x},t)\,{\cal P}(0)\,|0> $$
 $$ t \gg 1 \;\;\rightarrow { {|<PS|\,{\cal P}\,|0>|^2}
 \over {2\;M_{PS}a^3}}\;e^{-M_{PS} \cdot t}$$
\begin{equation}
 \equiv \zeta_{PP} \; e^{-M_{PS} \cdot t}
\label{ppcorr}
\end{equation}
${\cal P}(x)$ is an interpolating operator with the quantum numbers of 
a pseudoscalar meson and $A^L_\mu$ is the lattice axial vector current. 
Both ${\cal P}$ and $A^L_\mu$ are dimensionless.  
  From (\ref{apcorr}) and (\ref{ppcorr}) one has,

\begin{equation}
 a^{-1}<0|\,A^L_4\,|PS> \equiv af^L_{PS}\;aM_{PS} = 
{{\zeta_{AP} \sqrt{2aM_{PS}}} 
\over {\sqrt{\zeta_{PP}}} }  
\end{equation}
or,

\begin{equation}
a\,f^L_{PS} = \sqrt{{2 \over {aM_{PS}}}} \; { {\zeta_{AP}} \over
 {\sqrt{\zeta_{PP}}}}
\label{fpsl}
\end{equation}

\vspace{.1in}
\noindent
 In order to convert to a dimensionful continuum 
decay constant one needs the value of $a^{-1}$. This must come from a
separate calculation of some dimensionful quantity (the $\rho$ mass or the
light meson decay constant $f_\pi$  have been used in the past)
which is compared with experiment to determine $a^{-1}$.  Once $a^{-1}$ is 
specified one has,

$$
\sqrt{M_{PS}}\,f_{PS} = \sqrt{M_{PS}} \, f^L_{PS} \,Z_A $$

\begin{equation}
= Z_A\, \sqrt{2}  \; { {\zeta_{AP}} 
\over {\sqrt{\zeta_{PP}}}} \,a^{-3/2}
\label{fps}
\end{equation}

\vspace{.1in}
\noindent
$Z_A$ is the matching constant necessary to relate the lattice axial vector 
current to its continuum counter part.  It can be estimated using perturbation 
theory.  There have also been attempts to extract $Z_A$ nonperturbatively 
directly from simulations using Ward identities.

\vspace{.1in}

From eq.(\ref{fpsl}) and eq.(\ref{fps}) one sees that in order to obtain 
reliable results for $f_{PS}$  many factors have to be under good control.

\begin{itemize}

\item  $a^{-1}$

\item confidence that the ground state has been isolated when extracting
    $ { {\zeta_{AP}} \over {\sqrt{\zeta_{PP}}}}  $

\item  $Z_A$ including correct normalization of lattice fermions

\end{itemize}

\noindent
In addition a careful look at other systematic errors is necessary.

\begin{itemize}

\item  finite volume effects

\item extrapolation in the light quark mass

\item extrapolation in the heavy quark mass.  Many calculations 
     have been done in the 
      static limit or using Wilson fermions with masses smaller than
      $M_b$.

\item  finite lattice spacing errors.  This is particularly 
      important for Wilson fermions which have ${\cal O}(a)$ errors 
       unless an improved action is used.

\end{itemize}

\vspace{.2in}
\noindent
Many groups have been working on heavy meson decay constants, 
as can be seen by the following list.

\noindent
 UKQCD\cite{ukqcdfps}, BLS\cite{blsfps}, APE\cite{apefps,apestat},
 ELC\cite{elcfps}, 
  FNAL\cite{fnal}, PWCD\cite{pwcd}, Hashimoto\cite{hashimoto},
  LANL\cite{lanlps}, HEMCGC\cite{hemcgc}.

Fig.4. shows lattice results for the $D_s$ meson decay constant from 
a large number of groups.  Although $f_{D_s}$ is less crucial for extracting 
CKM matrix elements, nevertheless this is an important quantity for the 
lattice to calculate, since experimental values have now started to 
appear.  Experimental numbers from the CLEO\cite{cleofds}
 and WA75\cite{wa75} collaborations are 
also shown in Fig.4. and are consistent with the lattice numbers. 
At this conference another experimental result for $f_{D_s}$ was 
presented by the BES collaboration\cite{besfds}, 
$f_{D_s} = 434^{+153+35}_{-133-33}
 MeV$, which is still consistent with the other experimental numbers and 
with the lattice values.  Hopefully it will be possible to reduce the 
experimental errors, so that lattice calculations can be calibrated 
and better tested.  

\begin{figure}[htb]
\begin{center}
\setlength{\unitlength}{.006in}
\begin{picture}(500,400)(0,0)
\put(10,50){\line(0,1){330}}
\multiput(10,100)(0,50){6}{\line(1,0){3}}
\multiput(487,100)(0,50){6}{\line(1,0){3}}
\put(6,50){\makebox(0,0)[r]{100}}
\put(6,100){\makebox(0,0)[r]{150}}
\put(6,150){\makebox(0,0)[r]{200}}
\put(6,200){\makebox(0,0)[r]{250}}
\put(6,250){\makebox(0,0)[r]{300}}
\put(6,300){\makebox(0,0)[r]{350}}
\put(6,350){\makebox(0,0)[r]{400}}

\put(20,360){\makebox(0,0)[l]{$f_{D_s} [MeV]$}}

\put(10,50){\line(1,0){480}}
\put(10,380){\line(1,0){480}}
\put(490,50){\line(0,1){330}}

\put(5,10){\makebox(0,0)[l]{LANL}}
\put(50,30){\makebox(0,0)[l]{BLS}}
\put(100,10){\makebox(0,0)[l]{UKQCD}}
\put(175,30){\makebox(0,0)[l]{APE}}
\put(230,10){\makebox(0,0)[l]{ELC}}
\put(280,30){\makebox(0,0)[l]{HEMCGC}}
\put(375,10){\makebox(0,0)[l]{WA75}}
\put(440,30){\makebox(0,0)[l]{CLEO}}

\put(30,216){\circle*{5}}
\put(30,216){\line(0,1){15}}
\put(30,216){\line(0,-1){15}}

\put(75,180){\circle*{5}}
\put(75,180){\line(0,1){36}}
\put(75,180){\line(0,-1){36}}
\put(70,216){\line(1,0){10}}
\put(70,144){\line(1,0){10}}

\put(140,162){\circle*{5}}
\put(140,162){\line(0,1){46}}
\put(140,162){\line(0,-1){8}}
\put(135,208){\line(1,0){10}}
\put(135,154){\line(1,0){10}}

\put(200,190){\circle*{5}}
\put(200,190){\line(0,1){9}}
\put(200,190){\line(0,-1){9}}

\put(260,177){\circle*{5}}
\put(260,177){\line(0,1){15}}
\put(260,177){\line(0,-1){15}}

\put(325,238){\circle*{5}}
\put(325,238){\line(0,1){64}}
\put(325,238){\line(0,-1){64}}
\put(320,302){\line(1,0){10}}
\put(320,174){\line(1,0){10}}
\put(320,150){\makebox(0,0){$(n_f=2)$}}

\put(390,182){\makebox(0,0){x}}
\put(390,182){\line(0,1){69}}
\put(390,182){\line(0,-1){69}}
\put(385,251){\line(1,0){10}}
\put(385,113){\line(1,0){10}}
\put(420,90){\makebox(0,0){(experiment)}}

\put(455,294){\makebox(0,0){x}}
\put(455,294){\line(0,1){77}}
\put(455,294){\line(0,-1){77}}
\put(450,371){\line(1,0){10}}
\put(450,217){\line(1,0){10}}

\end{picture}
\end{center}
\label{fds}
\caption{Lattice and experimental results for the $D_s$-meson 
decay constant $f_{D_s}$.}
\end{figure}
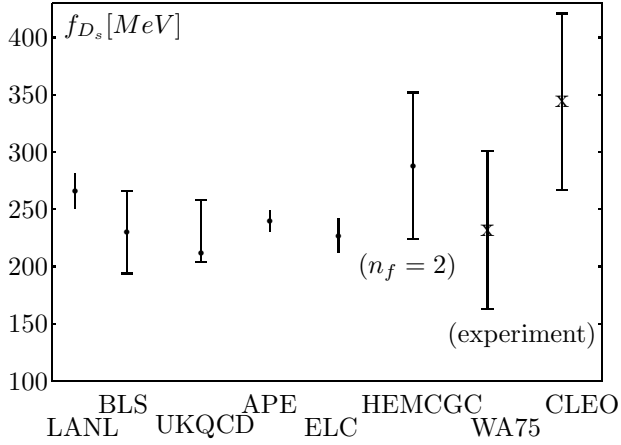

\begin{table}
\begin{center}
\begin{tabular}{l|c|c|c}
            & $f_D$[MeV]        & $f_{D_s}$[MeV]      & $f_{D_s}$/$f_D$\\
\hline\hline
&&& \\
  BLS[42] &$208 \pm 9 \pm 37$ &$230 \pm 7 \pm 35$ & $ 1.11 \pm .02 \pm .05$\\

&&& \\

  UKQCD[41]&$185^{+4+42}_{-3-7}$ & $ 212^{4+46}_{-4-7}$ &
 $1.18 \pm 0.02$ \\

&&& \\

PWCD[46]  & $170 \pm 30$    &   &  $1.09 \pm .02 \pm .05$  \\

&&& \\

APE[43]   & $218 \pm 9$    &  $240 \pm 9 $       &  $1.11 \pm.01$  \\

&&&  \\

ELC[44]  &  $210 \pm 15 $   &  $227 \pm 15$  &  $ 1.08 \pm .02$ \\

&&& \\

LANL[48]  &  $241 \pm 19 $  &   $266 \pm 15$ &  \\

&&& \\

HEMCGC[49] & $215 \pm 5 \pm 53$  &  $288 \pm 5 \pm 64$ & \\
($n_f = 2$)  &&& \\
\hline
&&&\\
Experiment &&& \\
$\;$WA75[52]  &  &  $232\pm 45 \pm 52$ & \\
&&&\\
$\;$CLEO[51] &  & $ 344 \pm 37 \pm 67 $ & \\
&&&\\
$\;$BES[53]  &  & $ 434^{+153+35}_{-133-33}$ &  \\
&&&\\
 \multicolumn{4}{c} {\bf Table 1. D Meson Decay Constants} \\
\end{tabular}
\end{center}
\label{tfd}
\end{table}

\vspace{.1in}
\noindent

\begin{table}
\begin{center}
\begin{tabular}{l|c|c|c}
            & $f_B$[MeV]        & $f_{B_s}$[MeV]      & $f_{B_s}$/$f_B$\\
\hline\hline
&&& \\
  BLS[42] &$187 \pm 10 \pm 37$ &$207 \pm 9 \pm 40$ & $ 1.11 \pm .02 \pm .05$\\

&&& \\

  UKQCD[41]&$160^{+6+53}_{-6-19}$ & $ 194^{+6+62}_{-5-9}$ &
 $1.22^{+4}_{-3}$ \\

&&& \\

PWCD[46]  & $180 \pm 50$    &   &       \\

&&& \\

ELC[44]  & $205 \pm 40$    &   & $1.08 \pm .06 $     \\

&&& \\

Hashimoto[47]  &  $171 \pm 22^{+19}_{-45} $   &    &  \\
(Nonrel.) &&&\\
&&&\\

FNAL[45] &  $188 \pm 23 ^{+34}_{-21}$  &   & $1.216 \pm .041 \pm .016$ \\
(Static)&&& \\
&&&\\

APE[50]  &  $290 \pm 15 \pm 45$    &     & $1.11 \pm .03$  \\
(Static) &&&\\
&&&\\
 \multicolumn{4}{c} {\bf Table 2. B Meson Decay Constants} \\
\end{tabular}
\end{center}
\label{tfb}
\end{table}

\vspace{.1in}
Tables 1. and 2. summarize
  values for D and B meson decay constants 
and for the ratios $f_{D_s}/f_D$ and $f_{B_s}/f_B$.  
At the moment it is not easy to compare the values for $f_B$ obtained 
by different groups, since each group has handled the sources of errors
listed above differently.
Even for gauge configurations with similar or identical parameters
 different values for $a^{-1}$ are being used, although most groups 
include $a^{-1}$ uncertainties in their systematic errors. 
  The last two entries in Table 2. use
static quarks ($M_Q \rightarrow \infty$) whereas the third from last
 entry uses 
nonrelativistic quarks. The first four groups worked with conventional 
Wilson fermions, normalized with varying degrees of sophistication in order 
to go smoothly to the large mass limit.  

\begin{figure}[hbt]
\epsfxsize=4.0cm
\centerline{\epsfbox{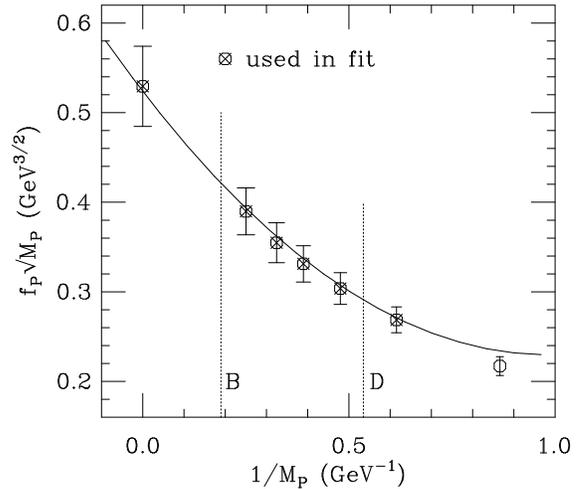}}
\caption{
Static and Wilson fermion results for pseudoscalar meson decay 
constants (from reference [42]).
}
\label{soni}
\end{figure}

The next couple of years should see
a vast improvement in these types of calculations as people learn from the 
current results and a consensus emerges on the best ways to proceed. 
Reference \cite{blsfps}
 was able to interpolate smoothly between D and B meson scales up to the 
static limit.  One figure from their paper is shown in \fref{soni}. 
Plotted is the combination given in eq.(\ref{fps}) rescaled by 
 $(1+\alpha_s\,ln(aM_P)/\pi)^{-1}$.  The fit is to an expression quadratic in 
$1/M_P$.  It was 
crucial for their good fit that they corrected for large $aM_Q$ errors 
coming from the use of Wilson fermions. 

\vspace{.3in}

\vspace{.1in}
\noindent
 These calculations are now being improved with higher statistics.
Looking elsewhere,
  methods that have been developed on heavy-heavy quarkonium 
systems such as the NRQCD and Fermilab Heavy Wilson approaches are starting 
to be applied to heavy-light D and B mesons.  With these approaches 
one can work directly at the B meson mass, and there is no need to 
extrapolate in the heavy quark mass.  One can expect a lot of activity 
here. 
 The issue of accurately 
determining the appropriate $a^{-1}$ in heavy-light systems as well as 
how to eliminate finite lattice spacing errors, will have to be 
resolved by all groups.  Only then can one hope to be able to provide 
experimentalists and phenomenologists with useful numbers with for instance 
$< 10$\% uncertainties.  
Nevertheless, at the present time we should feel encouraged that 
results obtained to date are all reasonably consistent with each other 
and that recent theoretical developments 
and better understanding of systematic errors point towards the possibility 
of significant improvements.

\vspace{.3in}

\section{Semileptonic D and B Meson Decays}

\subsection{$D \longrightarrow K , K^*$ Semileptonic Decays}

There is a growing body of results on semi-leptonic form factors 
coming from the lattice.  Table 3. gives a summary for $f_+$, 
$V$, $A_1$ and $A_2$ at $q^2 = 0$ for $D \longrightarrow K\,l\,\nu$ 
or $D \longrightarrow K^*\,l\,\nu$ decays.  Listed at the end is an 
average of experimental numbers.  One sees that the agreement among 
different lattice groups and between lattice predictions and experiment 
is good.  The generic form factor, $f(q^2)$, is evaluated at several 
different values of $q^2$ and a pole dominance ansatz $f(q^2) = 
f(0)/(1-q^2/M^2)$ is used to extract $f(0)$.  The relevant mass $M$ 
can either be fit simultaneously with $f(0)$ or obtained from an 
independent calculation of an appropriate correlation function. 
The next step in heavy to light meson decays is to go onto 
charmless B decays, $B \rightarrow \rho (\pi) \, l \nu$.  This will be 
a challenge for both experimentalists and lattice gauge theorists.  When 
successful, one will obtain a good handle on the CKM matrix element 
$V_{ub}$.

\begin{table}
\begin{center}
\begin{tabular}{c|c|c|c|c}
    & $f_+(0)$  & $V(0)$  & $A_1(0)$  &  $A_2(0)$ \\
\hline\hline
&&&& \\
  BKS[54] &.90(8)(21) & 1.43(45)(49)& .83(14)(28)  &.59(14)(24)  \\

&&&& \\

  UKQCD[55]&$.67^{+7}_{-8}$  & $.98^{+10}_{-13}$ & $.70^{+5}_{-10}$ &
   $.68^{+11}_{-17}$ \\

&&&& \\

APE[56]   &.72(9)  & 1.00(20)  &  .64(11)  & .46(34)  \\

&&&&  \\

ELC[57]  &.60(15)(7) & .86(24)  &  .64(16)  &.40(28)(4)  \\

&&&&  \\

LANL[58]   & .73(6)  &  1.24(8)  & .66(3)  & .45(19)  \\

&&&& \\
\hline
&&&& \\
\underline{Exp.}[59] &.77(4)  & 1.16(16)  & .61(5)  &  .45(9)  \\
(aver.) &&&&  \\

&&&& \\
 \multicolumn{5}{c}{\bf Table 3.  Semileptonic D  Decays} \\
\end{tabular}
\end{center}
\end{table}

\subsection{Isgur Wise Function}

There has been a lot of activity, both experimentally and theoretically 
on heavy meson to heavy meson semileptonic decays.  For instance, the 
decay $B \longrightarrow D^* \, l \, \nu$ has led to a determination 
of the CKM matrix element $|V_{cb}|$.  Heavy Quark Effective Theory (HQET) 
and the introduction of heavy quark spin and flavor symmetries and the 
Isgur-Wise function have had great impact on how to extract information from 
experimental data\cite{hqet1,hqet2}. 
 In the large $M_Q$ limit all the form factors, 
$h_+$, $h_-$, $h_V$, $h_{A_1}$, $h_{A_2}$, $h_{A_3}$ are related to the 
Isgur-Wise function $\xi(\omega)$ ($\omega = v \cdot v'$). 
  Experiments provide the combination $|V_{cb}|^2\,\eta_A\,
\hat{\xi}^2(\omega)$, 
where $\hat{\xi}$ means corrections to the heavy quark symmetry limit 
are included and $\eta_A$ includes short distance loop corrections.
 Since $\hat{\xi}(\omega = 1)$  and $\eta_A$ can be calculated, one can 
determine $|V_{cb}|$\cite{hqet3,sumr}. 

 Experiments are not done at $\omega = 1$ so one 
needs to extrapolate from $\omega > 1$ to $\omega = 1$.  An important 
quantity is the slope with which the Isgur Wise function approaches 
$\omega = 1$.

  $$ \rho^2 \equiv - \xi'|_{\omega = 1}  $$

\noindent
Several parametrizations of the Isgur-Wise function exist.  Using any 
one of them, close to $\omega=1$ one has

$$
  \xi(\omega) = 1 - \rho^2(\omega - 1) + {\cal O}((\omega-1)^2)  
$$

\noindent
Two lattice groups, BSS\cite{bss}
 and UKQCD\cite{ukqcdiw}, have made the first attempts to 
obtain the slope parameter $\rho^2$  from lattice QCD. 
 Combining this information 
with CLEO and ARGUS data they also determine $|V_{cb}|$.  Experimentalists 
(and theorists working with HQET) 
can, of course just fit their own data to extract $\rho^2$ and 
determine $|V_{cb}|$ (as they have done) without any input from the 
lattice. 
What lattice simulations can provide are more accurate 
data close to $\omega=1$ than is possible with current experiments and 
this should serve to guide fits to experimental data. It is also useful 
to be able to compare determinations of the slope parameter $\rho^2$ 
coming from a lattice QCD calculation with other theoretical estimates 
and with experiment.  Lattice 
calculations will eventually study the individual form factors
$h_+$, $h_{A_1}$ etc.  and investigate $1/M_Q$ corrections.  Some work 
in this direction  by the UKQCD collaboration has already appeared.

\vspace{.2in}
\noindent
The most accurate lattice results for the Isgur-Wise function come 
from studies of Pseudoscalar $\rightarrow$ Pseudoscalar decays, i.e.
using $h_+$.  BSS looks at the elastic $D \rightarrow D$ process, 
where current conservation helps to reduce systematic uncertainties.
  UKQCD considers $B \rightarrow D$ (and 
$B \rightarrow D^*$) decays.  Using the BSW parametrization of 
the Isgur-Wise function they obtain,

$$
 \rho^2  = \left\{ \begin{array}{ll}
                      1.4 \pm .2 \pm .4 & \;\; BSS  \\
                      0.9^{+2+4}_{-3-2} &  \;\;UKQCD
                      \end{array}
\right.
$$

\noindent
 Both groups find that the slope increases with the light quark mass $M_q$.
UKQCD finds, for
 instance that for 
$M_q \sim M_{strange}$ the slope increases from the above value to 
$\rho^2 = 1.2 \pm 0.2 ^{+.2}_{-.1}$.  
This $M_q$ dependence of $\rho^2$ agrees with QCD Sum Rule\cite{sumriw} and 
Quark Model\cite{qmiw} results reported at this conference, but disagrees with 
predictions from chiral perturbation theory\cite{cptiw}.

\vspace{.1in}
\noindent
From fits to their own and experimental data the two lattice groups find,

$$
 \sqrt{ \tau_B \over 1.53ps }|V_{cb}| = \left\{ \begin{array}{ll}
                      0.044 \pm 0.005 \pm 0.007 & \;\; BSS  \\
                      0.035^{+1+2+4}_{-1-2-1}  &  \;\;UKQCD
                      \end{array}
\right.
$$

\begin{figure}[htb]
\begin{center}
\leavevmode
\epsfysize=180pt
\epsfbox[20 30 620 600]{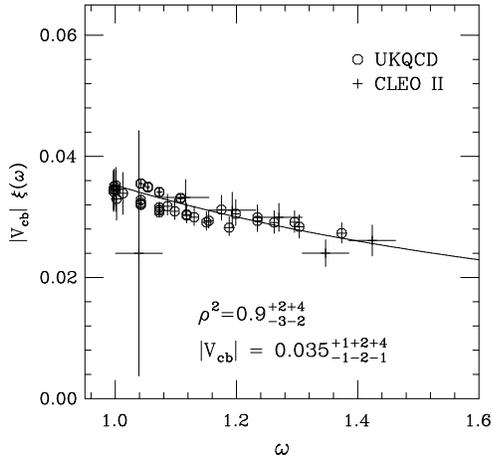}
\caption{UKQCD plot [65] to extract $|V_{cb}|$ using CLEO II Data.}
\label{vcbcleo}
\end{center}
\end{figure}

\noindent
  \Fref{vcbcleo} is a plot from 
the UKQCD collaboration, used to extract their value for $|V_{cb}|$.  It 
should be noted that, although radiative corrections at $\omega=1$ have 
been taken into account, exact heavy quark symmetry is assumed in relating 
the lattice simulations of the pseudoscalar $\rightarrow$ pseudoscalar 
process to experimental data.  UKQCD is currently also studying the 
decay $B \rightarrow D^*$, in order to obtain another estimate of $|V_{cb}|$.

\section{Radiative Penguin Decays, $B \longrightarrow K^*\, \gamma$}
The rare decay $B \rightarrow K^* \, \gamma$ was observed last year 
by the CLEO collaboration\cite{cleobkg}
and this year, at this conference, there was 
considerable excitment over their measurement of the inclusive 
$b \rightarrow s\,\gamma$ branching ratio\cite{cleobsg}.
The experimental numbers are,

$$
BR(b \rightarrow s \, \gamma) = (2.32 \pm .51 \pm .29 \pm .32) \times 10^{-4}
$$
$$
BR(B \rightarrow K^* \, \gamma) = (4.5 \pm 1.5 \pm 0.9 ) \times 10^{-5}
$$

\noindent
Lattice gauge theorists have also 
started to investigate rare B decays.  
For the process $B \rightarrow K^* \, \gamma$, 
what is required is a nonperturbative evaluation of the hadronic matrix 
element,

 $$ <K^*| \,\bar{s} \sigma_{\mu\nu} q^\nu b_R\,|B>  $$

\noindent
$q^\nu$ is the momentum of the emitted photon. This matrix element can 
be expressed in terms of three form factors, $T_1(q^2)$, $T_2(q^2)$ and 
$T_3(q^2)$.  On-shell one has,

\begin{equation}
 T_1(0) = T_2(0)
\label{t1t2}
\end{equation}

\noindent
and the ``hadronization ratio'' can be written as,

$$
R_{K^*}  \equiv Br(B \rightarrow K^* \gamma) /
                Br(b \rightarrow s \, \gamma) \qquad
$$
\begin{equation}
 = 4 [M_B/M_b]^3\,[1 - M_{K^*}^2/M_B^2]^3\,|T_1(0)|^2
\end{equation}

\vspace{.1in}
\noindent
Calculations are done at $q^2 > 0$ and to date also with masses smaller 
than the B meson mass.  Hence two important extrapolations, $q^2 \rightarrow 0$
 and $M_H \rightarrow M_B$ ($M_H$ is the mass of the initial decaying meson) 
are required.  Different groups are still exploring the best way to carry 
out these extrapolations. 

\vspace{.1in}
\noindent
 BHS\cite{bhs}
 extrapolates $T_2$ in $M_H$ up to $M_H = M_B$ using a form suggested 
by Heavy Quark Symmetry.  

\begin{equation}
\sqrt{M_H}T_2(q^2_{max}) = A_1 + A_2 /M_H
\end{equation}

\vspace{.1in}
\noindent
$q^2_{max} \equiv (M_H - M_{K^*})^2$.  They then use pole dominance to 
extrapolate from $q^2_{max}$ to $q^2=0$ and then obtain $T_1(0)$ from
eq.(\ref{t1t2}). 
 They were able to check for 
evidence in favor of  pole dominance over a small region in $q^2 > 0$,
 but not all the way to $q^2_{max}$ appropriate for $M_H = M_B$.  
They find,

\begin{equation}
T_1(0) = T_2(0) = 0.101 \pm 0.010 \pm 0.028
\end{equation}

\begin{equation}
R_{K^*} = (6.0 \pm 1.2 \pm 3.4)\%
\end{equation}

\vspace{.2in}
\noindent
UKQCD\cite{ukqcdbkg} uses pole dominance for $T_1$ to go to $q^2 = 0$ 
 and a modified scaling law, in which 
$T_1\,M_H^{3/2}$ is expanded in powers of $1/M_H$,  to extrapolate to $M_B$.  
They find,

\begin{equation}
 T_1(0) = 0.124^{+20}_{-18} \pm 0.022
\end{equation}

\begin{equation}
R_{K^*} = (8.8^{+2.8}_{-2.5} \pm 3.0 \pm 1.0)\%
\end{equation}

\vspace{.1in}
\noindent
The central values by BHS and UKQCD for $R_{K^*}$ 
appear to be  considerably smaller 
than what one would calculate from the CLEO results.  However within 
the large errors in both the lattice and the experimental numbers one still 
finds consistency. 

\vspace{.1in}
\noindent
At this conference preliminary results on $B \rightarrow K^*\,\gamma$ 
were presented by the APE collaboration\cite{apebkg}.
  Using pole dominance on $T_2$ 
they obtain $T_2(0) = 0.087(7)$.  However, using pole dominance on $T_1$ 
they find $T_1(0) = 0.20(7)$, a larger number.  It appears more work 
is needed to resolve the issue of how best to carry out the 
$q^2 \rightarrow 0$ and $M_H \rightarrow M_B$ extrapolations.
Lattice results for $B \rightarrow K^*\,\gamma$ 
should be viewed as very promising but still exploratory studies
at this time.  In the 
future in addition to improving the $B \rightarrow K^* \, \gamma$ calculations 
one would also want to go onto $B \rightarrow \rho\,\gamma$ 
processes.

\section{Summary}

These are exciting times for lattice gauge theory. More and more 
phenomenologically relevant quantities are now being studied on the 
lattice and  lattice results are starting to contribute towards 
testing and understanding the Standard Model
as we see it today.  It is also being used 
to investigate the Standard Model in extreme environments, such as 
QCD at finite temperatures and the Electroweak theory undergoing its
symmetry breaking phase transition.  There has been considerable 
progress in understanding how to correct for lattice artifact 
and other systematic errors and  
extract continuum physics. Precision calculations of $\alpha_s$, $M_b$ and 
$B_K$ have appeared.  
  There is optimism that significant improvements in the accuracy of 
many other lattice results are achievable.

\vspace{.3in}
\noindent
\underline{Acknowledgements}

\noindent
I wish to thank and congratulate the organizers of ICHEP94 for a very enjoyable
conference. 
I am also grateful to the many colleagues who gave valuable advice and 
assistance in putting together this review.  They include A.Abada, C.Bernard, 
S.Collins, C.Davies, A.El-Khadra, R.Gupta, K.Hornbostel, R.Kenway, G.Kilcup, 
A.Kronfeld, P.Lepage, P.Mackenzie, 
C.Morningstar, D.Richards, H.Shanahan, J.Sloan,
 A.Soni and A.Ukawa.

\noindent
The author's research is supported in part by the US Department of Energy,
DE-FG02-91ER40690.

%
%

\end{document}